\documentclass[11pt]{article}

\usepackage{graphicx}
\usepackage{cite}
\usepackage{mathpple}

\usepackage{geometry}
\begin{document}

{\small\noindent
\textbf{Preprint of:}\\
T. A. Nieminen, H. Rubinsztein-Dunlop and N. R. Heckenberg\\
``Calculation and optical measurement of laser trapping forces
on non-spherical particles''\\
\textit{Journal of Quantitative Spectroscopy and Radiative Transfer}
\textbf{70}, 627--637 (2001)
}

\vspace{6mm}

\hrulefill

\vspace{9mm}

\begin{center}

\LARGE
\textbf{Calculation and optical measurement of laser\\ trapping forces on
non-spherical particles}

\vspace{3mm}

\large
T. A. Nieminen, H. Rubinsztein-Dunlop, and N. R. Heckenberg

\vspace{3mm}

\normalsize
\textit{Centre for Laser Science, Department of Physics,\\ The University
of Queensland, Brisbane QLD 4072, Australia}

\texttt{timo@physics.uq.edu.au}

\end{center}

\begin{abstract}

Optical trapping, where microscopic particles are trapped and
manipulated by light is a powerful and widespread technique, with the
single-beam gradient trap (also known as optical tweezers) in use for a
large number of biological and other applications.

The forces and torques acting on a trapped particle result from the
transfer of momentum and angular momentum from the trapping beam to the
particle.

Despite the apparent simplicity of a laser trap, with a single
particle in a single beam, exact calculation of the optical forces and
torques acting on particles is difficult. Calculations can
be performed using approximate methods, but are only applicable
within their ranges of validity, such as for particles much larger
than, or much smaller than, the trapping wavelength, and for spherical
isotropic particles.

This leaves unfortunate gaps, since wavelength-scale particles are
of great practical interest because they are readily and strongly
trapped and are used to probe interesting microscopic and macroscopic
phenomena, and non-spherical or anisotropic particles, biological, crystalline, or
other, due to their frequent occurance in nature, and the possibility of
rotating such objects or controlling or sensing their orientation.

The systematic application of electromagnetic scattering theory can
provide a general theory of laser trapping, and render results missing
from existing theory. We present here calculations of force and torque
on a trapped particle obtained from this theory and discuss the
possible applications, including the optical measurement of the force
and torque.

\vspace{3mm}

\end{abstract}

\section{Introduction}
\label{introduction}

\begin{figure}[!htbp]
\centerline{\includegraphics[width=3in]{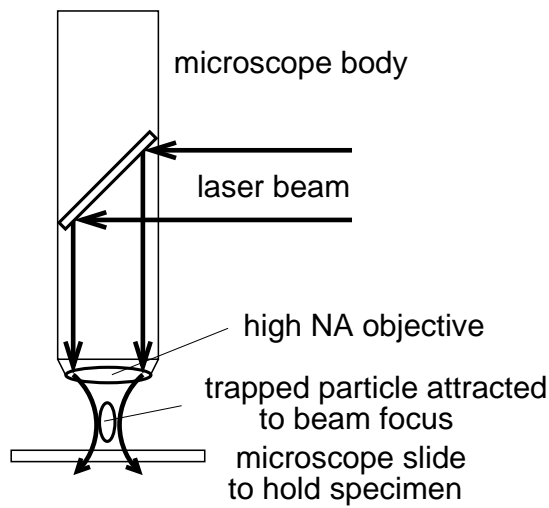}}
\caption{Schematic diagram of a typical optical tweezers setup}
\label{tweezersfig}
\end{figure}

Optical trapping, which is the trapping and manipulation of microscopic
particles by a focussed laser beam or beams, is a widely used and
powerful tool. The most common optical trap, the single-beam gradient
trap, commonly called {\em optical tweezers}, consists of a laser beam
strongly focussed by a lens, typically a high-numerical aperture
microscope objective, with the same microscope being used to view the
trapped particles (see fig. \ref{tweezersfig}) \cite{svoboda1994}. The
trapped particle is usually in a liquid medium, on a microscope slide.
Commonly used laser sources employed for trapping range from He-Ne
lasers, through Ar ion and semiconductor lasers to TiS and NdYAG lasers.
Varying laser powers are used in a broad range of applications of
optical tweezers - from just a few milliwatts to hundreds of mW. For
most of the lasers used, when the beam is passed through the objective
lens, the focal spot of the trapping beam is of the order of a micron.
The trapped objects can vary in size from hundreds of nanometres to
hundreds of microns.

Although simple trapping and manipulation are sufficient for many
applications, the use of optical trapping for quantitative research into
physical, chemical, and biological processes, typically using a
laser-trapped particle as a probe, requires accurate calculation of the
optical forces and torques acting within the trap. The approximate
methods commonly used for such calculations may prove inadequate for
many applications \cite{grier1997}. An accurate quantitative theory of
optical trapping not only allows such calculations to be performed, but
also greatly extends the usefulness of techniques such as optical force
and torque measurement and optical particle characterisation.

The concept of optical trapping is based on a gradient force causing
small particles to be attracted to regions of high intensity in a
tightly focussed laser beam \cite{ashkin1972}. Other optical forces, due
to absorption, reflection, and scattering are termed scattering forces.
Both the gradient and scattering forces result from the transfer of
momentum from the trapping beam to the particle. Optical torques can
also be produced by the transfer of angular momentum from the beam,
which can result from birefringence, or particle shape, or absorption of
a beam carrying non-zero angular momentum
\cite{friese1998a,luo2000,he1995,friese1996}.

Although the optical forces and torques within a trap result in a
straightforward manner from the change in momentum and angular momentum
due to scattering processes, exact calculation is difficult, and a
number of approximations are usually made. Approximate calculations use
geometric optics for large particles (radius $a > 5 \lambda$), or assume
that a small particle ($a < \lambda/2$) acts as a Rayleigh scatterer or
a point-like polarisable particle. This leaves a large range of particle
sizes without adequate results; an unfortunate gap since particles of
size comparable to the trapping wavelength are of great practical
interest because they can be readily and strongly trapped, and can be
used to probe interesting microscopic and mesoscopic phenomena. Reecnt
theoretical efforts have individually eliminated some of the
deficiencies due to the various approximations used, but there still
exists no general correct theory
\cite{ren1996,wohland1996,farsund1996,neto2000}.

The lack of suitable theory is even more acute when non-spherical or
anisoropic particles are considered. Non-spherical paricles are of
particular interest due to their suitability for use as microscopic
probes, and their frequent occurence in nature, for example, biological
structures and crystals are usually non-spherical, and are often
anisotropic. The possibility of rotating or controlling the orientation
of such particles greatly extends the range of manipulation possible
within an optical trap, introducing new applications, such as, for
example, the investigation of microscopic rotational dynamics
\cite{nieminen2000}.

These theoretical deficiencies can be overcome by considering the
scattering processes responsible for optical trapping. There exists a
well-developed body of work on electromagentic scattering which can be
applied to laser trapping in order to determine the scattered fields,
from which, in turn, the optical forces and torques can be found.

There are some examples in the literature of attempts to develop a
general theory of optical trapping. These are typically restricted to
a limited range of particle types and sizes. The systematic application
of scattering theory can eliminate these limitation, giving a general
theory correct for all particles compositions, including transparent,
conductive, absorbing, etc., all particle sizes and shapes, and for
arbitrary trapping fields.

\section{Light scattering in an optical trap}
\label{scattering}

An optical trap in most ways presents a simple electromagnetic
scattering problem, with usually only a single particle in a single
orientation in the trap at any one time. The major problems are the
representation of the beam, and the possibility that the particle is
such that calculations will be difficult even for a single particle.
A number of the available techniques require that the trapping beam
(i.e. the incident field) be represented as a plane wave spectrum or in
terms of vector spherical wavefunctions (VSWFs). The trapping beam is
usually a strongly focussed (i.e. non-paraxial) Gaussian beam. An
immediate problem is that the standard representations of Gaussian beams
do not actually satisfy the Maxwell equations, leading to some
difficulty in finding a plane wave or VSWF spectrum, though
satisfactory methods exist \cite{barton1989,doicu1997,gouesbet1999}.
Additionally, the trapping beam cannot actually be Gaussian, but will
have been truncated at some point in the optical system. Non-Gaussian
beams are also used for trapping, for example, Laguerre-Gaussian
``donut'' beams \cite{heckenberg1999}.

We can note that while the existence of a wide variety of techniques
for the calculation of scattering is indicative of the lack of a
universally superior method, each technique has its own particular
advantages and disadvantages, and we can find a usable or even ideal
method for any particular case. A brief survey of commonly used
methods follows, focussing on application to trapping problems. Computer
codes implementing many of these methods are available
\cite{wriedtweb,flatauweb}.

In general, the solution of an electromagnetic scattering problem
requires the solution of the Maxwell equations. Some geometries
yield relatively simple solutions, more general cases require direct
numerical solution of the Maxwell equations. The best known analytical
method is Mie theory, restricted to scattering by a homogeneous
isotropic sphere illuminated by a plane wave. Extensions of Mie theory,
including the use of spheroidal expansions instead of spherical
coordinates allow a broader, but still very limited, range of
applicability. In general, numerical methods must still be used to
obtain the final solutions in these cases \cite{mishchenko1999}.

Closely related to Mie theory is a family of numerical techniques where
the incident and scattered fields are expressed in terms of VSWFs, and
the expansion coefficients of the scattered field are found by the
boundary conditions at the surface of the scatterer. These methods
include the point matching method \cite{wriedt1998} and the T-matrix
method \cite{mishchenko1998,mishchenko1999b}. The T-matrix method is
widely used, computationally efficient for axisymmetric particles, and a
number of computer codes implementing this technique are freely
available \cite{wriedtweb,flatauweb,mishchenkoweb}. The T-matrix method
is of particular interest, since, for a given scatterer and illuminating
wavelength, the T-matrix only needs to be computed once, and can then
be used for repeated calculations. A surface integral over the particle
must be computed, but in the case of a rotationally symmetric particle,
this integral reduces to one dimension \cite{tsang1985}.

These surface-based techniques are, in their simple forms, restricted to
homogeneous particles, though extensions to layered particles exist
\cite{peterson1974}. Other surface methods include the generalised
multipole technique \cite{hafner1990} and the method of moments
\cite{wriedt1998,mishchenko1999,bancroft1996}.

If the scattering particle is such that techniques such as those above
cannot be used, there are a number of general techniques, in principle
usable for any scattering problem. In general, these methods are
computationally intensive \cite{wriedt1998,mishchenko1999}.

Since the Maxwell equations are a set of differential equations, finite
difference or finite element methods can be used. The finite difference
time domain method (FDTD) is widely used in computational
electromagnetics, and can be applied to scattering problems
\cite{taflove1995}. A discrete grid of points in space is set up, and
the fields at successive time steps are calculated. Since the
discretisation in space must be much smaller than the wavelength
($\approx \lambda/20$), and a correspondingly small time step must be
used, only a relatively small volume can be used for the calculation, so
the boundary conditions at the edge of the computational volume must be
carefully chosen, and if the far-field is desired, it must be found from
the near-field via a suitable transformation. Finite element
methods (FEM) can also be used, again using spatial discretisation to
obtain a numerical solution to the system of differential equations
\cite{volakis1998,white2000}. Both of these methods a conceptually
simple, and can represent a particle of any shape or composition, but
are computationally intensive and require special care with the
boundary conditions.

An alternative method is the disrete dipole approximation (DDA) where
the particle itself is divided into small volumes, each of which can be
treated as a simple dipole with a polarisability depending on the
composition of the particle. An initial guess of the final field is
iteratively improved until convergence is obtained
\cite{draine1994,draine1999}. Computer codes implementing DDA are
publicly available \cite{wriedtweb,flatauweb,draineweb}.

We can note that the characterisation of a laser trap will require
repeated calculations with the same particle in different positions and
orientations within the trap. Thus, the T-matrix method is attractive,
since the T-matrix need only be found once. The T-matrix method requires
that the incident beam be represented in terms of VSWFs; this can be
done directly, or via an intermediate plane wave expansion. The need for
repeated calculations makes the general methods (such as FDTD, FEM and
DDA) less attractive since the entire calculation must be repeated if
the incident beam is changed. In these cases, the total calculation
required can be minimised by representing the trapping beam by a plane
wave spectrum, and calculating the scattering for the different angles
of plane wave illumination. This represents no greater computational
effort than calculating the scattering for all orientations of the
particle at a single location within the trap, and gives results that
can be used to find the sccattering at all points and all orientations.

\section{Calculation of forces}
\label{calculation}

Once the scattering has been calculated, the resulting force and torque
can be found from the field, by integrating around the particle. The
momentum of an electromagnetic field is given by \cite{jackson1999}

\begin{equation}
\mathbf{P}_{\mathrm{field}} = \epsilon_0 \int_V
   \mathbf{E} \times \mathbf{B} d^3x
\end{equation}

and the angular momentum by

\begin{equation}
\mathbf{L}_{\mathrm{field}} = \epsilon_0 \int_V
   \mathbf{x} \times (\mathbf{E} \times \mathbf{B}) d^3x
\end{equation}

If we have the far-field scattered field, which is an outgoing radiation
field, we can choose a spherical surface for the integration, and the
momentum and angular momentum fluxes are normal to the surface. Then,
the field can be represented by the electric field alone, with two
orthogonal components, with complex amplitudes $E_\theta$ and $E_\phi$,
$\mathbf{E}_\theta = E_\theta \hat{\theta}$ and
$\mathbf{E}_\phi = E_\phi \hat{\phi}$, tangential to the surface of
integration. The rate of transfer of linear and angular momentum per
unit area can be found using \cite{nieminen2000}

\begin{equation}
\mathbf{P}_{\mathrm{flux}} = \frac{\epsilon_0}{2}
   (E_\theta E_\theta^\star + E_\phi E_\phi^\star) \hat{\mathbf{n}}
\label{momentumeqn}
\end{equation}

and

\begin{equation}
\mathbf{L}_{\mathrm{flux}} = \frac{i c\epsilon_0}{2\omega}
   (E_\theta E_\phi^\star - E_\theta^\star E_\phi) \hat{\mathbf{n}}
\label{angularmomentumeqn}
\end{equation}

where $\hat{\mathbf{n}}$ is the unit vector normal to the surface of
integration. Equation (\ref{momentumeqn}) can be readily recognised
as the familiar result of the Poynting vector divided by $c$.

From the conservation of momentum and angular momentum, the
integrated momentum and angular momentum fluxes give the force and
torque acting on the trapped particle.

If the discrete dipole approximation (DDA) is used to calculate the
scattering, the particle is represented by a number of dipoles, within a
known field (after the scattering calculations have been performed).
Therefore, the force and torque on each dipole can be found, giving the
force and torque on the entire particle, and incidentally, the stresses
within the particle. This allows the force and torque to be found
without needing to integrate around the particle \cite{hoekstra2000}.

The basic methods can be summarised as:

\begin{itemize}
\item If the T-matrix method can be used:
     \begin{enumerate}
     \item calculate T-matrix
     \item find VSWF representation of beam at the desired position
           within the trap, for the desired orientation
     \item find scattered field (in terms of VSWFs)
     \item integrate around particle
     \end{enumerate}
\item If a general method must be used
     \begin{enumerate}
     \item calculate plane wave scattering by the particle for all
           angles of incidence
     \item find plane wave spectrum of trapping beam at the desired
           position
     \item find the scattering for each spectral component, and combine
           to find the total scattered field
     \item integrate around particle
     \end{enumerate}
\item If DDA is used
     \begin{enumerate}
     \item find scattering at desired point
     \item find forces and torques on all dipoles, and combine to find
           the total force
     \end{enumerate}
\end{itemize}

\begin{figure}[!htbp]
\centerline{\includegraphics[width=4.5in]{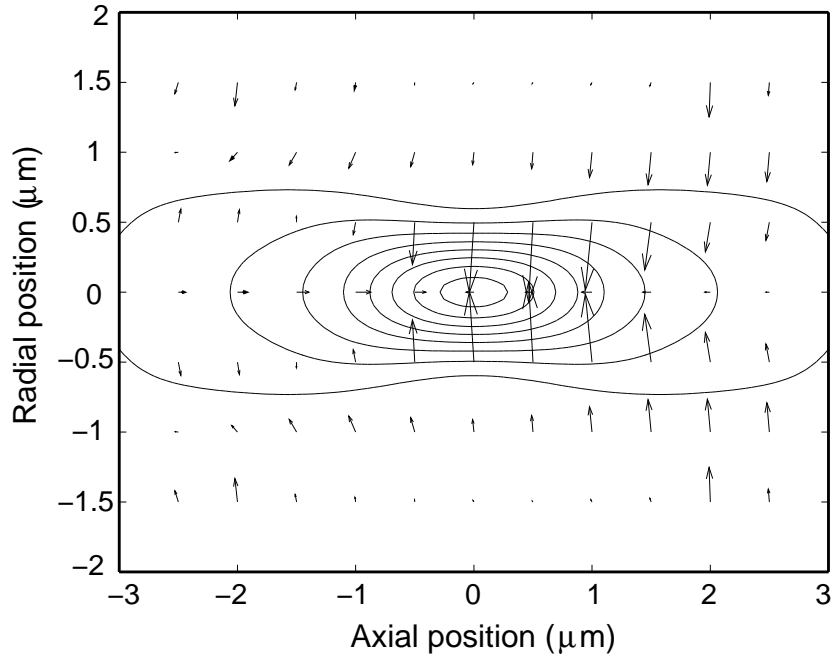}}
\caption{Optical force on spheroid}
\label{forcequiverplot}
\end{figure}

Some sample force calculations are presented in figure
\ref{forcequiverplot}. The trapped particle is a glass prolate spheroid
($n = 1.5 + 0.02i$), with $a = b = 0.5\mu\mathrm{m}$ and $c =
1.0\mu\mathrm{m}$, trapped in water ($n = 1.33$) by a Gaussian beam of
free space wavelength $1064 \mathrm{nm}$, focussed to a spot width of
$1 \mu\mathrm{m}$. The T-matrix code developed by Mishchenko
\cite{mishchenkoweb} was used to find the T-matrix an amplitude matrix.
The trapping beam was decomposed into a plane wave spectrum consisting of
97 components. The size parameter of the particle is $4.95$. The
wavelength and beam profile and particle size, shape and composition
were chosen to best model the most typical situation occurring in the
experimental realisation of the technique.

Figure \ref{forcequiverplot} shows the forces exerted by the trapping
beam, which is propagating from the left to the right, focussed to a
small spot in the centre. The magnitude of the force is proportional to
the length of the arrows, and the direction is the same as the direction
of the arrows. The contours in the background are equal-intensity
contours for the trapping beam. The trapping beam intensity falls of
with increasing radial distance due to the narrow width of the beam. The
intensity falls off axially due to the spreading of the beam. The strong
radial trapping forces are clearly visible, and it can be seen that the
forces are approximately normal to the intensity contours. Since the
radial intensity gradient is so large, the radial forces are the largest
optical forces acting on the particle. In the regions of low intensity
(at the top and bottom of the middle of the graph), the forces are very
small.

\section{Optical measurement of forces and torques}
\label{measurement}

The forces and torques acting on the trapped particle result from the
scattering within the trap, and can be calculated if the scattered
fields are known. So far, we have considered techniques for the
calculation of the the scattered field. In many cases, this will not be
possible, for example, if our particle is a biological
structure within a living cell, with unknown optical properties, or a
particle with known optical properties at an unknown position within the
trap.

If the scattered radiation field can be experimentally measured,
equations (\ref{momentumeqn}) and (\ref{angularmomentumeqn}) can be used
to find the force and torque by integrating over a surface around the
particle. In principle, the required measurement of the scattered field
is possible, but the application of this will be limited by a number of
practical considerations, namely:
\begin{enumerate}
\item The normal method for measuring scattered light from trapped
particles is to measure either the back-scattered light
\cite{friese1996b}, or the forward scattered light
\cite{ghislain1993,pralle1999}. Back-scattering measurements usually
record the total power received by the detector and do not give the
directional resolution required for measuring the scattered field
(which, in any case, would be virtually impossible to do with any
accuracy after the return passage of the back-scattered light through
the microscope optics). Forward scattering measurements usually use low
resolution detectors, such as quadrant detectors, and do not make use of
all the possible measurable information. It is generally impossible to
collect all of the scattered light due to the spatial limitations
imposed by a typical optical tweezers setup. To collect more than the
forward scattered light emergent from the bottom of the trap will be
very difficult. The placement of the detector(s) is limited by the
design of the laser trap. Since the trapping cell, contained a
microscope slide, will restrict collection of side-scattered light, and
the focussing lens (the objective) will restrict the collection of
back-scattered light, the detector must be placed below the trapping
cell (see fig. \ref{ccdplacement}). Since the detector must be capable
of spatial resolution, and must be large enough to measure the
forward-scattered light that passes through the trapping cell, a CCD
array is ideal.

\begin{figure}[!htbp]
\centerline{\includegraphics[width=3.5in]{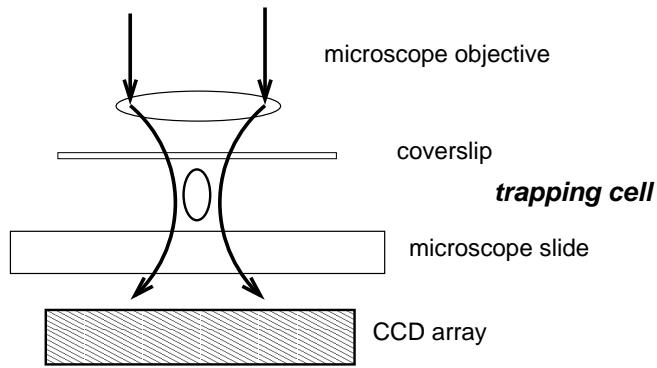}}
\caption{Optical trap with a CCD array to measure scattered light}
\label{ccdplacement}
\end{figure}
A CCD array, by itself, is not sufficient to measure the field; only the
intensity will be measured, and it will still be necessary to measure
the state of polarisation. This can be done by incorporating a polariser
acting as an analyser (see fig. \ref{detector}).
\begin{figure}[!htbp]
\centerline{\includegraphics[width=3.5in]{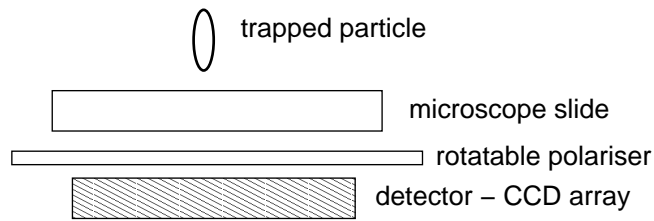}}
\caption{Using a polariser and CCD array to measure the polarisation
state of light scattered by a trapped particle}
\label{detector}
\end{figure}
\item The bottom of the trap will reflect and refract the scattered
light. Since the scattered light is initially in a higher refractive
index medium (the fluid within the trapping cell and the glass
microscope slide), some of this light will be totally internally
reflected, and will be unmeasurable (see fig. \ref{reflection}). For
typical values of refractive index, the maximum scattering angle for
which the light can be measured is approximately $45^\circ$. Reflection
can also occur at interfaces internal to the trapping cell, although
these will be smaller since the refractive index differences will be
relatively small. The reflections that do occur will depend on the
polarisation of the light. If too much light is scattered at large
angles (and therefore unmeasurable), it will not be possible to
accurately determine the optical force. Being restricted to measuring
forward-scattered light will also make it difficult to measure torques
other than that acting about the beam axis. The partial reflection of
light that exits the trapping cell and is measured can, and in the
interests of accuracy, should be compensated for.
\begin{figure}[!htbp]
\centerline{\includegraphics[width=3.5in]{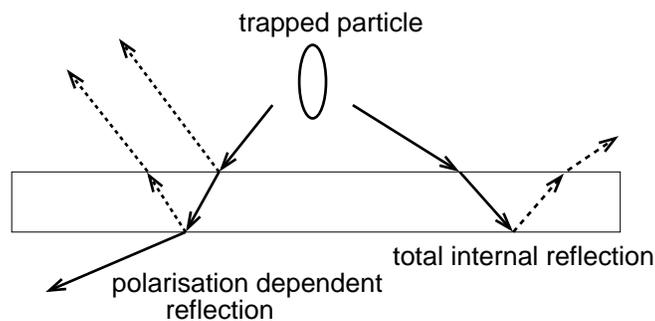}}
\caption{Total internal limits the maximum angle at which
light can be scattered and still exit the trapping cell}
\label{reflection}
\end{figure}
\item The resolution and size of the detector will limit the
measurements.
\end{enumerate}

The restriction of the measurable scattering to the forward-scattered
light limits the applicability of the technique. It remains useful,
however, in the two cases of most interest, one case being organisms and
biological structures, the second probe particles.

Organisms and biological structures in a trap will typically have
unknown optical properties, and often have complex shapes. Thus, it will
be necessary to measure the scattering since calculation will be
impossible. Since such particles are optically soft (with a relative
refractive index $m \approx 1$), very little light will be scattered at
large angles, and the majority of the scattered light can be measured,
and the force and torque acting on the particle determined with
reasonable accuracy.

If the particle in the trap is not optically soft, with a large
refractive index difference compared with the surrounding medium, it
will not be possible to collect enough of the scattered light to
accurately determine the force and torque in all cases. If, however, the
particle in question is a probe particle of known size, shape, and
optical properties, but at an unknown position within the trap, the
portion of the scattered light that is measured can be used to determine
what the total scattering pattern must be. Thus, the position of the
probe particle within the trap, and the optical force and torque acting
on it, can be determined. In this way, the external non-optical forces
acting on the probe particle can be determined \cite{nieminen2000aos}.

A special case of the use of a trapped particle as a probe is the
rotating probe particle, which can be used to measure the viscosity of
a fluid, colloid, or suspension on a microscopic scale. The rotating
particle will typically be birefringent, and will remain in the centre
of the trap. The change in polarisation of the light on scattering will
cause the particle to rotate. Both the optical torque acting on the
particle and its rotation speed can be measured \cite{nieminen2000}.

\section{Conclusion}
\label{conclusion}

Approaching laser trapping as a scattering problem allows the
calculation of forces and torques using electromagnetic scattering
theory. Such calculations can be performed for all types of particles:
transparent, abosorbing, conductive, reflective, anisotropic, complex
shapes, etc. For particles for which efficient computational methods
(such as the T-matrix method) can be used, calculations are fast and
relatively simple, and can be performed on readily available PCs.

This means that the gap in previous calculations, where no adequate
results were available for particles comparable in size to the trapping
wavelength, can be closed.

Optical measurement of forces and torques acting on particles within the
trap can also be performed, by measuring the scattered light. Apart from
being free of the usual calibration difficulties for force measurement
in optical traps, it can be used where traditional force measurements
are impossible, such as measurement of the forces acting on structures
within living cells.

Measurement of the scattered light from a known probe particle, coupled
with calculation of the scattering in different positions of the trap,
allows measurement of the position of, and force acting on, the probe
particle. Thus, external non-optical forces can be determined. A special
case of this, where the probe is a rotating trapped particle, is
particularly simple, and the rotation speed and optical torque can be
simultaneously measured.

\end{document}